\documentclass[twocolumn,showpacs,preprintnumbers,amsmath,amssymb]{revtex4}

\usepackage{graphicx}% Include figure files
\usepackage{dcolumn}% Align table columns on decimal point

\begin{document}

\title{Two kinds of spin precession modes in diluted magnetic semiconductors}

\author {M.~Vladimirova, S. Cronenberger, P.~Barate,  D. Scalbert}
\affiliation{Groupe d'Etude des Semi-conducteurs, UMR 5650
CNRS-Universit\'{e} Montpellier 2, Place Eug\`{e}ne Bataillon, 34095
Montpellier Cedex, France} %%
\author {F. J. Teran}
\affiliation{Departamento de Física de Materiales, Universidad
Aut\`{o}noma de Madrid, 28049 Madrid, Spain}
%%    Information for the second author

\author {A. P. Dmitriev}
\affiliation{A. F. Ioffe Institute, 26, Polytechnicheskaya, 194026,
St-Petersburg, Russia,  and  \\ Groupe d'Etude des Semi-conducteurs,
UMR 5650 CNRS-Universit\'{e} Montpellier 2, Place Eug\`{e}ne
Bataillon, 34095 Montpellier Cedex, France}

\begin{abstract}
%
%We demonstrate the collective character of the spin exitations in
%n-type CdMnTe quantum wells.
%
Time-resolved Kerr rotation experiments show that two kinds of
 spin modes exist in diluted magnetic semiconductors: (i)
coupled electron-magnetic ion spin excitations  and (ii) excitations
of magnetic ion spin subsystem, which are decoupled from electron
spins.
The latter modes exhibit much longer spin coherence time and require
a description, which goes beyond the mean field approximation.
\end{abstract}

 \maketitle

%**************************************************************************
Diluted magnetic semiconductors (DMS) are well recognized as
spin-model systems, and are considered as particularly attractive
components for hybrid spintronic devices \cite {Kikkawa}.
In a DMS, a small fraction of the cations is  randomly substituted
by magnetic ions, i. e. manganese (Mn).
Their most important properties, such as ferromagnetic phase
transition and spin excitations dynamics are controlled by the
interaction between the spins of  itinerant carriers and the spins
of the magnetic ions.

At low magnetic field, the exchange interaction strongly amplify the
spin precession frequency of the carriers.
At higher fields the exchange field saturates, and  the spin
precession frequency gets controlled  by Zeeman splitting.
This allows to vary the detuning between carrier spins precession
frequency $\omega_e$ and Mn spins precession frequency $\omega_m$.
Under resonance conditions, $\omega_e=\omega_m$,  electron and ions
spin excitations are strongly coupled, which results in the
formation of two collective spin precession modes \cite{Teran},
\cite {KonigMD}.
They correspond to the in-phase and anti-phase spin precession of
the average electron and ion spins (Fig.1a, b).
Their frequencies are given by the well known formula for the two
coupled oscillators :
\begin {equation}
\omega_{\pm}=\frac{1}{2} (\omega_e+\omega_{m})\pm\frac{1}{2}
\sqrt{(\omega_e-\omega_{m})^2+4\delta^2} \label {omegapm}
\end {equation}
where $\delta$ is the interaction energy (Fig. 1d, solid lines).
At resonance field the key characteristics of the coupled modes are
the avoided crossing and the identical decoherence time.
The later is  limited by the electron spin relaxation, which is much
faster than the spin relaxation of the ions.

In this Letter  we show, that another kind of spin excitations
exists in DMS.
Despite the strong resonant coupling of the Mn spins to the
carriers, these modes appear to be pure Mn spin excitations, which
do not involve electron spin.
We don't observe neither the frequency shift with respect to the
bare Mn spin precession (Fig.1d, dashed line), nor acceleration of
the magnetic ions spin relaxation under resonant conditions.
In order to interpret this astonishing fact we go beyond the
standard description in terms of average electron and ion spins
interaction.
The decoupled modes appear to have the  distributions of the
out-of-equilibrium components of Mn spins such that the exchange
field that they create on the electrons is zero, so that electron
spin stays at equilibrium (Fig.1c).

The samples that we study are two n-type CdMnTe quantum wells with
two-dimensional concentration of magnetic ions $n_m=2.4 \cdot
10^{13}$ cm$^{-2}$.
Electron concentrations are $n_e=0.7 \cdot 10^{11}$ cm$^{-2}$
(sample 1), $n_e=2.2 \cdot 10^{11}$ cm$^{-2}$ (sample 2).
The well width is $w=$100 $\AA$ \cite {Cox}.
\begin{figure}[]
\includegraphics[width=1.1\columnwidth]{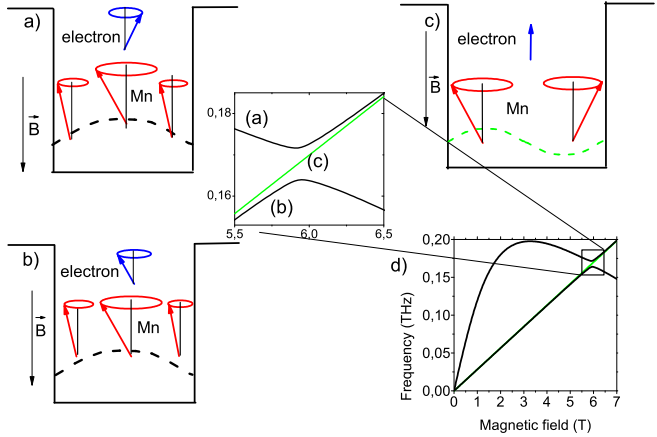}
\caption{\label{fig1} (Color online) Schematic representation of the
mean field spin precession modes (a,b) and the long-living mode,
where the magnetic ions are decoupled from the carriers (c). Dashed
lines map the distribution of the transverse spin component for the
magnetic ions. Arrows are the snapshot of  the precessing spin
vectors of magnetic ions (long arrows shown in the center and at the
interfaces) and electron (short arrows). Field dependence of the
mode frequencies are shown in (d). The inset enlarges the resonance
region.}
\end{figure}
The spin excitations are identified using all-optical spin resonance
 technique \cite {KikkawaScience}.
The  100 fs pulses produced by a mode-locked Al$_{2}$O$_{3}$:Ti
laser with a repetition rate of 82~MHz are spectrally filtered to
obtain $\sim$1.5 ps pulses tuned to the heavy hole exciton
transition. %
The laser light beam is separated into the pump (200 $\mu$W) and
probe (100 $\mu$W)
The polarization of the pump is modulated  between left and right
helicities at 50 kHz, in addition to the intensity modulation of
both pump and probe beams.
Both beams are focused on the $200$ $\mu$m diameter spot on the
surface of the sample.
The pump-induced rotation of the linear polarization of the
reflected probe (Kerr rotation) is measured as a function of delay
between pump and probe pulses.
It is usually assumed to be proportional to the spin polarization in
the direction of the light.
The sample is placed in the cryostat at $T=2$ K under magnetic field
$B\sim 6$ T  in the plane of the sample (along $z$-axis).
\begin{figure}[]
\includegraphics[width=1.1\columnwidth]{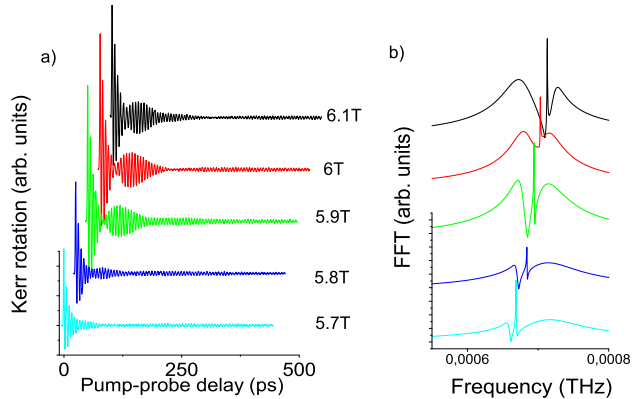}
\caption{\label{fig2} (Color online)  Time-resolved Kerr rotation
signal under magnetic field from 5.5 T to 6.1 T  (a) and the
corresponding Fourier spectra (b).}
\end{figure}

Let us consider what happens when DMS quantum well is excited by the
pump pulse.
Prior to the  excitation the magnetic ions are strongly spin
polarized in $z$ direction.
Circularly polarized pump pulses create in the quantum well about
$10^{10}$ cm$^{-2}$ electrons and holes spin polarized normal to the
plane, in the $x$-direction.
The hole spin is locked in this direction, because the $g$-factor of
the hole is vanishing in the direction perpendicular to the quantum
well axis.
Nevertheless, the hole spin acts on the magnetic ion spins as an
effective magnetic field, until the hole spin is fully relaxed
($\tau_h\sim$5 ps).
It coherently rotates the spins of the magnetic ions away from the
$z$-direction and initiates the precession of the Mn spins around
the total magnetic field, created by photocreated carriers and the
external field \cite {CrookerPRL}.
The electron spin is also expected to precess around the field,
created by the hole spins, Mn spins and the external field.
Thus, because Kerr rotation signal is proportional to the
$x$-component of the magnetization in the sample, we expect to
measure the dynamics of the optically generated transverse spin
component resulting from the ensemble of the interacting spin
subsystems.

Fig.2 shows a set of Kerr rotation measurements under magnetic
fields from 5.7 to 6.1 T for the sample 1 (a) and the corresponding
Fourier spectra obtained after the fitting procedure (b) \cite
{notesample1}.
We identify in these curves four different components.
The first one is the exponential decay of the signal during first
$\sim$5 ps after the excitation.
It is due to the hole spin relaxation.
Two other components are the damped cosines, the corresponding decay
times and frequencies strongly vary from 5.7 to 6.1T.
The frequencies are plotted as a function of magnetic field in Fig.
3a (circles and squares).
One can see a clear anticrossing behavior.
These are the collective modes discussed in Refs. \cite {Teran,
KonigMD}, where precessing electron an Mn spins are strongly
coupled.
Fig. 3b shows the  relaxation times of the collective modes.
\begin{figure}[]
\includegraphics[width=1.1\columnwidth]{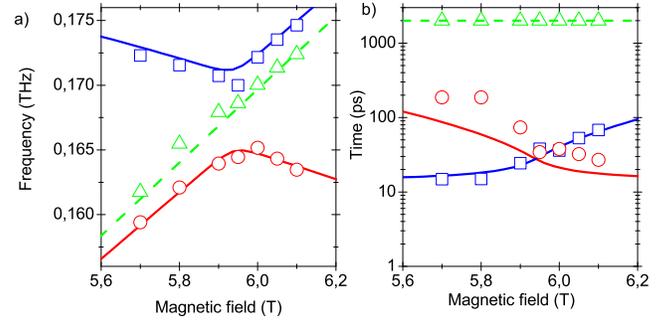}
\caption{\label{fig3} (Color online)  Spin precession frequencies
(a) and spin coherence times (b) extracted from the data shown in
Fig. 2a. Symbols show the experimental data, lines are the the best
fit obtained using Eq. (1) assuming electron spin polarization of
70\%. Soled lines indicate the collective modes, dashed lines show
Mn spin excitations  decoupled from electrons.}
\end{figure}%
As it can be expected for the two coupled oscillators with very
different quality factors (like electron and magnetic ion spins),
the spin relaxation time of the Mn-like mode (circles) is
dramatically reduced  in the vicinity of the resonance.

However, careful analysis of the data  shows that the correct
description of the signal imposes taking into account a forth
oscillating component (Fig. 2).
It is related to another, unexpected spin precession mode.
It has the longest relaxation time, which could not be precisely
determined, but estimated to be at least of 2000 ps.
The field dependence of this long-living mode frequency is linear,
and is given by the $g$-factor of 2.02, which corresponds precisely
to the Mn spin  $g$-factor (Fig. 3, triangles) \cite {Crooker},
\cite {Akimoto}.
It is unlikely due to the spatial regions in the well where the
carriers are absent.
Indeed, in both samples the electron gas is strongly degenerate and
the electrons are fully delocalized in the plane of the quantum well
\cite {Cox}.
The existence of the long-living spin excitations is further
corroborated by the resistively detected electron paramagnetic
resonance experiments of Ref. \cite {Teran}, where an additional
feature at the Mn spin resonance frequency was detected, though not
discussed in the paper.

We develop here the model for the electron gas coupled to the
magnetic ions, which allows for the description of both the complex
spectrum of the spin excitations in this system and the dephasing of
the modes.
It is based on the standard Hamiltonian:
\begin {eqnarray}
\nonumber
%\begin{multiline}
H=H_{0}+V+g_e\mu_B\sum_\beta(\mathbf{S}_\beta\mathbf B)
+g_{m}\mu_B\sum_k(\mathbf{J}_k\mathbf B)- \\
{\raggedleft
\alpha\sum_{\beta,k}\delta(\mathbf{r}_\beta-\mathbf{R}_k)(\mathbf{S}_\beta\mathbf{J}_k).~~~~~~~~~~~~
}
%\end{multiline}
\label{Hgeneral}
\end {eqnarray}
Here $H_{0}$ defines the kinetic and potential  energy of the
electrons confined in the quantum well, $V$ is the impurity
scattering potential, including the potential of magnetic ions.
$\mathbf{S}_\beta$, $\mathbf{J}_k$ are electron and magnetic ion
spin operators, respectively.
The  $g$-factors of electrons and Mn are $g_e=-1.5$ and $g_m=2$,
respectively, $\mu_B$ is the Bohr magneton \cite {Sirenko}.
The last term in the Hamiltonian (\ref {Hgeneral}) describes the
exchange interaction of the ferromagnetic type between the electrons
and Mn spins, $\alpha=1.5\times 10^{-23}$ eV/cm$^3$.
Vectors $\mathbf{R}_k$ and $\mathbf{r}_\beta$ denote the positions
of the ions and the electrons, respectively.

We limit our consideration to  the strong  fields $B\sim5-6$ T, so
that at equilibrium the magnetic ions are almost fully polarized in
the direction opposite to the magnetic field and $J_z=-5/2$.
The polarization of the electrons is mainly determined by the strong
exchange field created by the spin polarized ions and thus oriented
in the same direction.
%
%The magnitude of the electron spin polarization  amounts to 70\% in
%the sample 2 (20\% in sample 1), since the electron gas is strongly
%degenerate.
%
%Indeed, the Fermi energy is equal to $2$ meV for the sample 1 and
%$5$ meV for the sample 2, both much higher than the thermal energy.
%
%Note also, that the thermal energy is much smaller than the energy
%separating the Fermi level from the second electronic subband, so
%that only one electron subband in the quantum well is occupied by
%electrons.

Starting from the Hamiltonian (\ref {Hgeneral}) we  derive the
equations of motion for the electron and Mn spin operators and
average them over the initial state, where both electron and ion
spins are slightly tilted with respect to their orientation at
equilibrium.
Bearing in mind that the non-equilibrium components of all the spins
are small, with respect to their equilibrium components, we
linearize the equations of motion about the equilibrium spin values
$S_z$ and $J_z$.
In order to simplify the calculations we assume the continuous
density of the magnetic ions and replace the summations over their
coordinates by the integration according to $\sum_k \rightarrow
(n_m/w)\int d^3 \mathbf{R}$, where $w$ is the width of the quantum
well.
This assumption seems reasonable because the concentration of the
ions is much larger than the concentration of the electrons, so that
a volume defined by the electron de Broglie wavelength contains
$\sim500$ ions.
Besides, we suppose that the magnetic ion spins are excited
homogeneously in the plane of the sample. %
We end up with two equations of motion for the average
non-equilibrium spin components normalized  by the corresponding
equilibrium spin values, $\sigma(t)=(S_x(t)+iS_y(t))/S_z$  for
electrons and  $j(x,t)=(J_x(x,t)+iJ_y(x,t))/J_z$ for magnetic ions.
Looking for the solutions in the form $\sigma(t)=\sigma_0e^{i\omega
t}$ and $j(x,t)=j_0(x)e^{i\omega t}$ we obtain the equation on the
eigenfrequencies $\omega$ and eigenvectors $(\sigma_0, j_0(x))$:
\begin{eqnarray}
\omega \sigma_0=\Omega_e\sigma_0 -\Delta \bar{j}_0 +i\frac
{\sigma_0}{\tau_e}~~~~~~~~~~~~~~~~~~~~~~~~~~~~~~~~~~~~ \label{el}
 \\
\omega j_0(x)=(\Omega_m+K \kappa(x))j_0(x)-Kw\chi^2(x)
\sigma_0~~~~~~~~~~~~\label{mn}
\end{eqnarray}
where $\chi(x)$ is the electron wavefunction at the lowest quantized
state, $\Omega_e=g_e\mu_B B/\hbar+\Delta$, $\Delta=-\alpha
n_mJ_z/w\hbar$, $\Omega_m=g_m\mu_B B/\hbar$, $K=-\alpha n_e
S_z/w\hbar$, $\kappa(x)=w\chi^2(x)-1$, $\bar{j}_0=\int_0^w
\chi^2(x)j_0(x)dx$ \cite{NoteKonigMD}.
Here we account for the short electron spin relaxation time $\tau_e$
by introducing phenomenologically the relaxation term in Eq.
(\ref{el}).

The first terms in the righthand part of the Eqs. (\ref{el}) and
(\ref{mn}) describe the precession of the electron and ion spins.
Both corresponding frequencies contain two contributions.
The first one comes from the Zeeman effect, while the second
accounts for the exchange field created by the equilibrium spin
polarization of the ions and electrons, respectively.
For electrons, the precession frequency is mainly determined by the
exchange field.
This field depends on the equilibrium polarization of the ions,
which is given by the Brillouin function, saturating at $B\sim5$ T.
Therefore, because $g_e<0$ in CdMnTe, the electron spin precession
frequency decreases with the magnetic field above $5$ T.
In contrast, $x$-dependent exchange contribution in the spin
precession of the ions is very small compared to the Zeeman
contribution (in our sample three orders of magnitude smaller).
While it can be neglected in the eigen frequencies calculation
procedure, it appears to be important for the calculation of the
relaxation times.

Now it is convenient to come back to the discrete form of  Eqs.
(\ref{el}) and (\ref{mn}).
The two-dimensional layer decomposes naturally into N components,
where N is the number of the atomic layers in the quantum well.
Denoting by $x_n$ the coordinates of these layers we obtain:
\begin{eqnarray}
\omega \sigma_0= \Omega_e\sigma_0- \Delta \bar{j}_0
+i\frac{\sigma_0}{\tau_e} ~~~~~~~~~~~~~~~~~~~~~~~~~~~~~~~~~~
 \label{el1}
 \\
\omega j_n=(\Omega_m +K \kappa_n)j_n-
 K w \chi_n^2 \sigma_0 ~~~~~~~~~~~~~~~~~~~~~~~~
 \label{mn1}
\end{eqnarray}
where $\kappa_n=\kappa(x_n)$, $\chi_n^2=\chi^2(x_n)$, and
$\bar{j}_0=w/N\sum_{n=1}^N \chi_n^2$.
Let us first neglect the small term $K \kappa_n$ in the righthand
part of the Eq. (\ref {mn1}).
%
%
%This correction accounts for the variation  of the individual ion
%spin precession frequency $\delta\omega \sim K$ and is mainly due to
%the difference between the probability to find an electron in the
%center of the two-dimensional structure and at the interface.
%
The resulting equations can be easily solved analytically.
%
%Indeed, multiplying  Eq. \ref{mn1} by $\chi_n^2$, dividing by $N$
%and making summation  over $n$ we obtain a system of two linear
%equations for $\sigma_0$ and $\bar{j}_0$.
%
%The solvability condition for this system implies two solutions
%given by ( \ref{omegapm}), with $\omega_e=\Omega_e-i\tau_e$ and
%$\omega_m=\Omega_m$, $\delta^2=\eta\Delta K$,  $\eta= w
%\int_0^w\chi^4(x)dx$ \cite {infbarr}.
%
They have two solutions for  $\sigma_0$ and $\bar{j}_0$ with
frequencies $\omega_{\pm}$ given by (\ref{omegapm}), where
$\omega_e=\Omega_e+i/\tau_e$ and $\omega_m=\Omega_m$,
$\delta^2=\eta\Delta K$,  $\eta= w \int_0^w\chi^4(x)dx$ \cite
{infbarr}.
These are the collective modes discussed in Ref. (\cite {Teran,
KonigMD})  \cite {noteMD1}.
Our model allows also for the estimation of their relaxation times.
At resonance the relaxation times of both collective modes are equal
$\tau=2\tau_e$.
The calculated frequencies and the decay times of the collective
modes are shown in Fig.3 by solid lines, they are in good agreement
with the experimental data.
Note, that the spin polarization degree of the electron gas is the
only fitting parameter in this calculation ( 70\% in sample 1), the
electron spin decay time $\tau_e=15$ ps being extracted from the fit
of the Kerr rotation signal at low magnetic fields, i.e. far from
the resonance.

However, the possible spin excitations are not exhausted by the
above collective modes.
Indeed, in the approximation made above ($K \kappa_n\sim0$) Eqs.
(\ref {el1}) and (\ref {mn1})  are satisfied if the electron spin is
not excited, so that $\sigma_0=0$, and at the same time $\bar
{j}_0=0$ (Fig. 1c).
These conditions describe the excitations for which the total
exchange field created by  the non-equilibrium components of the ion
spins are zero, and therefore they do not affect the equilibrium
electron spin.
The corresponding eigenfrequency  $\omega=\Omega_m$ fits the
experimental results (dashed line in Fig. 3a).
The number of such degenerate modes is $N-1$.
The relaxation time associated with each of these modes is infinite
\cite {Mnspinrel}, as far as the correction $K\kappa_n$ is neglected
in Eq. (\ref {mn1}).

Let us now take into account the correction $K\kappa_n$.
The Eqs. (\ref {el1}) and (\ref {mn1})  can still be solved
analytically.
As a result we obtain that the collective modes frequencies
$\omega_{\pm}$ and relaxation times are only slightly affected by
this term.
Its main effect is the lifting of the degeneracy of the N-1
"everlasting" modes, leading to the broadening of the corresponding
spin resonance $\delta \omega \sim K$.
One can show, that approximately half of these modes remains
decoupled from the electron spin, and therefore does not decay.
The remaining modes weakly interact with the electrons and thus
acquire a relaxation time of the order of $\tau_e \Delta/K$.
In our experiments $\tau_e \Delta/K>>1/K$.
Therefore we expect the decay  of the Kerr rotation signal from the
long-living spin excitations at the frequency $\Omega_m$ after a
characteristic time
 $\tau_m \sim 1/K \sim 1$ ns.
This is a good order of magnitude (dashed line in Fig.3b), while it
is much shorter than any homogeneous relaxation time \cite {Konig}.
Thus, at resonance, Mn spin dephasing due to non-uniform
distribution of the electron density in the growth direction appears
to be the dominant spin decay mechanism.

Finally, we discuss the possible source of  the electron spin
relaxation  $\tau_e\sim$15 ps.
This value is considerably shorter than the theoretical predictions
\cite {Semenov}.
We suggest, that electron spin dephasing is governed by the
fluctuations of the exchange field created by the ions on the
 scale of the electron wavelength $\lambda$ and, therefore,
it is intimately related with the fluctuations of the concentration
of the ions on this scale.
This effect is missing in the Eqs. (\ref {el})-(\ref{mn1}), which
are written assuming  continuous and homogeneous distribution of the
ion density.
The estimation for our sample leads to $1/\tau_e\sim \ (\alpha/
\hbar \lambda^2)\sqrt{n_m\lambda^2}\sim (\alpha/\hbar )\sqrt{n_m
n_e}\sim 10^{11}$ s$^{-1}$ in good agreement with the experimental
value.
Thus, we assert that the dephasing of both electron and Mn spin
excitations is governed by the spatial inhomogeneity of the
corresponding exchange fields, and is  stronger for the electrons
$\tau_e<<\tau_m$ because their concentration  is much smaller than
Mn concentration.

In conclusion, we have identified  a new kind of spin excitations in
DMS quantum wells.
These excitations do not involve electrons, which are  at
equilibrium, while the Mn spins precess as if they were not affected
by the exchange interaction.
Neither their  frequency, nor their relaxation time is affected by
the strong exchange interaction with electron spin.
The formation of the collective modes discovered by Teran {\it at
al} \cite {Teran} is shown to be responsible for the dramatic
reduction of the Mn spin relaxation time under resonant conditions.
We are able to interpret the ensemble of the experimental
observations in the framework of a simple semiclassical model going
beyond the mean field approximation.
%

%

%**************************************************************************
%
%**************************************************************************
\begin{acknowledgements}
We acknowledge helpful discussions with M. I. Dyakonov and the
support from the Marie-Curie RTN project 503677 "Clermont2". A.P.D.
acknowledge the support from RFBR, Russian Scientific School, and
programmes of RAS.
\end{acknowledgements}
\vskip 1truecm


\begin{thebibliography}{99}
\bibitem{notesample1} The results are very similar for the sample 2.
\bibitem{NoteKonigMD} These equations can be obtained by the
minimization of the affective action calculated in Ref. \cite
{KonigMD}, Eq. (8).
\bibitem{infbarr} In the approximation of the infinite barriers the
electron wavefuction is $\chi(x)=(\sqrt{2/w}) sin(\pi x /w)$ and
$\eta=3/2$.
\bibitem{noteMD1} Factor $\eta$ is, however, absent in the
expression for $\delta^2$ from Ref.(\cite{KonigMD}) and the role of
$\bar{j}_0$ is played by the average spin of the ions.
\bibitem{Mnspinrel} The coherence time of each "everlasting" mode is
anyway limited by the relaxation of the individual Mn spins, mainly
on the lattice phonons.
\end{thebibliography}
\end{document}